\documentclass[12pt]{iopart}

\begin{document}

\title[On the possibility of variation of the fundamental constants 
of physics]{On the possibility of variation of the fundamental 
constants of physics in the static universe}

\author{V. Jonauskas\footnote[1]{e-mail: jvaldas@itpa.lt}}

\address{State Institute of Theoretical Physics and Astronomy,
A. Go\v stauto 12, 2600 Vilnius, Lithuania}

\begin{abstract}
A variation of fundamental constants of physics is proposed in a frame of 
static universe. It is shown when the velocity of light increases 
(decreases) the Planck's constant increases (decreases) and mass of bodies 
decreases (increases). This variation of constants leads to the variation of 
dimensions of bodies and the energy levels of atoms, but a fine structure 
constant remains unaltered. 
\end{abstract}

\maketitle

\section{Introduction}

In 1938 Dirac \cite{dir38} developed a cosmology based on a remarkable
numerical coincidence among the fundamental quantities of physics. He
proposed the gravitational constant changes with time only. There were many
efforts to register this experimentally. Other scientists were varying
different physical constants (the electron mass, the light velocity, etc.)
to explain Large Numbers in physics or 
Hubble law \cite{bell77,man76,troit87}.

We formulate our question another way: how fundamental constants and how
many of them could be changed if the universe around us remained the same at
all points of space and at all times. The synchronous variation of a number
of fundamental constants is considered in the work \cite{troit87}. The
static cosmological model has been made, based on the hypothesis of
continuous variation (decrease) of the light velocity in vacuum. The author 
\cite{troit87} gets the variation of the fine structure constant, and this
variation may be observed for an interval of 10 years.

Our work is based on a ''perfect cosmological principle'': the universe is
spatially isotropic and homogeneous and it looks the same at all times \cite
{bon48}. We also suggest that the universe expands with time. 
In our work we are
not making any suggestion about the expansion rate of the universe,
described by the Hubble's constant. The classical steady state model
confirms the expansion must always occur at the same rate ${H}_{0}$
(Hubble's constant) as we measure today. 
The greatest shortcomings of static model
are an inability to describe microwave background radiation 
and the abundance of light elements in the universe.
Troitskii explains the microwave background radiation by thermal radiation
of galaxies \cite{troit96}. Burbidge and Hoyle conclude that all
of the chemical elements and microwave background radiation were
produced from hydrogen burning in stars \cite{burb98}.

The synchronous variation of fundamental constants is proposed in 
\cite{jon98}. Assumptions of the constancy of the fine structure constant, 
the Coulomb and Newton forces and electric charge are made. Here we start 
with a concept that all bodies including an atom,  galaxies and the universe 
expand with time. Thus the universe is static for an observer. We assume 
that the velocity of light is constant for any local observer. One of the 
ways to make this assumption valid when dimensions of bodies increase is to 
suggest that the velocity of light increases at the same scale as expansion 
of bodies. We are getting that the Planck's constant increases and masses of
bodies decreases with time.

\section{Assumptions}

An expansion of the universe  is accepted to explain 
redshift of a light from remote celestial bodies. 
A wavelength ${{{\lambda }_{0}}}$ observed here
at time ${t_{0}}$ is related to a wavelength ${{{\lambda }_{1}}}$ emitted at
time ${t_{1}}$ by

\begin{equation}
\frac{{{\lambda }_{0}}}{{{\lambda }_{1}}}=\frac{R({t_{0}})}{R({t_{1}})},
\label{e1}
\end{equation}
where R(t) is a cosmic scale factor at time $t$. If the universe is
expanding, then R( ${t_{0}}$) $>$ R( ${t_{1}}$), and (1) gives a
red shift, while if the universe is contracting, then R( ${t_{0}}$) 
$<$ R( ${t_{1}}$), and (\ref{e1}) gives a blue shift.

We are making first our assumption: dimensions of all bodies varies with
time at the same scale as the cosmic scale factor. Distances between bodies
drawn on a rubber balloon expand as the balloon inflates and the dimensions
of the bodies on the balloon increase at the same scale as the distances
between them. The dimensions of an atom also have to expand. The expansion
of bodies should be understood as averaged through the universe.

Our tools of the measurements are related to bodies and we are not able to
check this altering of space and dimensions. So, the universe should obey a
''perfect cosmological principal'' \cite{bon48}: it looks the same not only
at all points and in all directions, but also at all times. It is a steady
state model of the universe, but without the necessity of continuous
creation of matter. However, it is also the static universe for an observer
because he is not able to register this expansion.

When the length unit increases and the velocity of light and time rate
remain unaltered the observer will register that the velocity of light
decreases with time. By this, we mean that the length unit increases but it
stays the length unit for the observer at all times. We are not able to
observe its increasing in the other way but the result of decreasing of the
velocity of light. We will save Einstein's assumption of the constancy of
the velocity of light by making another assumption, the assumption that the
velocity of light increases from time ${t_{1}}$ to time ${t_{0}}$ ( ${t_{1}}$
$<$ ${t_{0}}$ ) by the factor:

\begin{equation}
{\Omega ({t_{1}}-{t_{0}})=\frac{{{\lambda }_{0}}}{{{\lambda }_{1.}}}}
\label{e2}
\end{equation}
Meaning, $\Omega $ $>$ 1. Further, we will use $\Omega $ instead
of $\Omega $( ${t_{1}}-{t_{0}}$). So we have supposed that dimensions of
bodies are less when the velocity of light is smaller and a clock rate is
the same. Therefore, observers are unable to register an altering of $c$.
These assumptions can be written as

\begin{equation}
{c_{0}}={c_{1}}\Omega ,  \label{e3}
\end{equation}
and

\begin{equation}
{l_{0}}={l_{1}}\Omega .  \label{e4}
\end{equation}
At time ${t_{0}}$ ( ${t_{1}}$), ${c_{0}}$ and ${l_{0}}$ ( ${c_{1}}$, ${l_{1}}
$) are the velocity of light and the length, respectively. From (\ref{e4})
we conclude that the velocity ${v_{0}}$ of the free body at time ${t_{0}}$
could be expressed by the velocity ${v_{1}}$ of the same body at time ${t_{1}%
}$ as

\begin{equation}
{v_{0}}={v_{1}}\Omega .  \label{e5}
\end{equation}
It is important to note that the body has not been acted on by any forces
when it was moving from time ${t_{1}}$ to time ${t_{0}}$. An observer will
not register this  velocity alteration since the dimensions alter at the
same scale and the time rate remains unchanged.

\section{Increasing of Planck's constant}

The assumption of the constancy of clock rate in the universe at all time
means that forces acting between bodies do not change. A time derivative of
momentum is equal to the force. So the momentum of a free body is not
altering. It could be written as a product of body mass $m$ and velocity 
$v$. So, we get from equation (\ref{e5}) the relationship:

\begin{equation}
{m_{0}}={m_{1}/}\Omega ,  \label{e6}
\end{equation}
where ${m_{0}}$ ( ${m_{1}}$) is the body mass at time ${t_{0}}$ ( ${t_{1}}$%
). The mass of the body is decreasing with time. Atoms are a constitution of
each body. The dimension of the atoms could be expressed in units of Bohr
radius which is equal to:

\begin{equation}
a_{0}=\frac{h^{2}4\pi {\epsilon }_{0}}{m_{e}e^{2}},  \label{e7}
\end{equation}
where ${{{\epsilon }_{0}}={{10}^{7}/}4\pi {c^{2}}}$ is the permittivity of
empty space, ${{m_{e}}}$ is the electron mass, $e$ is the electron charge
and $h$ is the Planck's constant. The variation of ${{{\epsilon }_{0}}}$
depends on the variation of c. According to (\ref{e3}), (\ref{e4}), (\ref{e6}%
) and (\ref{e7}) we could express Planck's constant ${h_{0}}$ at time ${t_{0}%
}$ through Planck's constant ${h_{1}}$ at time ${t_{1}}$ when the electrical
charge remains unchanged:

\begin{equation}
{h_{0}}={h_{1}}\Omega ,  \label{e8}
\end{equation}
meaning, Planck's constant increases with time as the velocity of light.

The variation of these fundamental constants leads to the variation of
bodies dimensions. This corresponds to changes of space geometry. When the
variation of constants depends on the space and time coordinates, we have
curved spacetime. This curving of spacetime is caused by the changes of the
linear dimensions of bodies. If the linear dimensions of an observer are
less in some area than the linear dimensions of another observer in the same
area then this area will be larger for the first observer and he will need 
more time to cross this area. So, we have curved spacetime. Meanwhile, 
dimensions of bodies change in our curved spacetime. Thus, our spacetime 
differs from the accepted model of curved spacetime, when dimensions of 
bodies remain unchanged but only space and time alters.

\section{Variation of some physical quantities}

The suggestion of unaltering time means that the gravitational and
electromagnetical forces remain unchanged in any situation. When Newton force

\begin{equation}
F=\gamma \frac{m_{1}m_{2}}{r^{2}}  \label{e9}
\end{equation}
remains unchanged, from equations (\ref{e4}) and (\ref{e6}) we get

\begin{equation}
{\gamma }_{0}={\gamma }_{1}{\Omega }^{4}.  \label{e10}
\end{equation}

The gravitational constant $\gamma $ increases with time. 
A ratio of the gravitational potential

\begin{equation}
\varphi =\gamma \frac{m}{r}  \label{e11}
\end{equation}
with a square of the light velocity is constant. The Coulomb force

\begin{equation}
F=\frac{1}{4\pi {\epsilon }_{0}}\frac{{q_{1}}{q_{2}}}{r^{2}}  \label{e12}
\end{equation}
between two charges ${q_{1}}$ and ${q_{2}}$ separated by a distance $r$
remains unchanged according to our assumptions made above. The dimensionless
constant (fine structure constant)

\begin{equation}
\alpha =\frac{1}{4\pi {{\epsilon }_{0}}}\frac{e^{2}}{hc}  \label{e13}
\end{equation}
remains unaltered after variations of ${{{\epsilon }_{0}}}$, c and $h$ as we
have proposed. In the same manner, we could get the variation of other
electromagnetical quantities: 
\begin{eqnarray}
\overrightarrow{E}_{0} &=&\overrightarrow{E}_{1},  \label{e14} \\
\overrightarrow{B}_{0} &=&\overrightarrow{B}_{1}/\Omega ,  \nonumber \\
\rho _{0} &=&\rho _{1}/\Omega ^{3},  \nonumber \\
\overrightarrow{j}_{0} &=&\overrightarrow{j}_{1}/\Omega ^{2},  \nonumber
\end{eqnarray}
and for $\overrightarrow{{\nabla }}$ operator

\begin{equation}
\overrightarrow{{\nabla }}_{0}=\frac{1}{\Omega }\overrightarrow{{\nabla }}%
_{1}.  \label{e15}
\end{equation}
Here $\overrightarrow{{E}}$ and $\overrightarrow{{B}}$ are the electric and
magnetic field vectors, respectively; $\rho $ is the charge density and $%
\overrightarrow{{j}}$ is the current density.

The above mentioned variation of constants if applied to the Maxwell
equations remains these
equations without changes, when $\Omega $ is constant, or dependence of $%
\Omega $ on time or space coordinates could be neglected. In the same way we
may learn how we could vary other fundamental constants or quantities of
physics.

\section{Hubble law}

Our proposed method of varying fundamental constants of physics corresponds
to changes of energetical distances between levels in atoms. For example,
energies of allowed states of the hydrogen atom are

\begin{equation}
E_{n}=-\left( \frac{m_{e}e^{4}}{{8\,\epsilon _{0}^{2}\,{h^{2}}}}\,\right) 
\frac{1}{n^{2}}.  \label{e17}
\end{equation}
After varying ${\epsilon }_{0}$, $h$ and ${{m_{e}}}$ we get

\begin{equation}
E_{n_{0}}=E{_{n_{1}}}\Omega .  \label{e18}
\end{equation}
This means that when $\Omega $ $>$ 1, the corresponding levels of
hydrogen atoms were higher, and the distances between them were smaller in
the past.

So, a photon emitted at time ${t_{1}}$ has lower energy than the photon
emitted at time ${t_{0}}$ ( ${t_{1}}$ $<$ ${t_{0}}$) by the same
atom. An atomic spectrum from remote bodies is shifted to the red side
because of the finite value of $c$. We should note that all physical
quantities with same dimension varies in the same way during synchronous
variation of constants. Thus, all lines (not only described by equation 
(\ref{e17})) of distant objects are redshifted.

A time measurement based on  transitions between levels in atoms shows the
increase of the time rate when the light velocity increases. Meanwhile, a
dynamical clock (for example, the period of revolution of the Earth about 
the Sun) rate does not depend on the light velocity variation.

It means that the light velocity measurement by the material length standard
and the quantum clock will show the decrease of the light velocity with
time. It is important to note the light velocity increases but the
measurement shows the decrease. This happens since the quantum time rate
increases when the light velocity and length standard increases too.
Troitskii \cite{troit87} gets that the light velocity measured in this way
remains unchanged.

The length of meter definition \cite{coh87} as a distance which the light
covers in 1/299792458 seconds does not have meaning when it is not mentioned
type of clocks. The measurements of time by dynamical and quantum clocks
will show different length standard.

\section{Conclusions}

As we see, assumptions of the bodies expansion with time and the constancy
of the light velocity for a local observer 
leads to a variation of  other 
fundamental constants. Accepted cosmological theory forbids the expansion of 
bodies such as atoms, the Earth, the Sun and so on. The expansion of all 
material bodies is forbidden primarily because there are no criterions by 
which the expansion of the universe could be measured. We perceive this 
separation as very strange because all these objects could be characterized 
by their dimensions. Therefore, in our cosmological model all bodies feel 
expansion because of the variation of the light velocity, Planck's constant, 
and bodies mass. A consequence of this variation is the steady state model
without the creation of mass. The variation of the fundamental constants
gives the Hubble law. Also, it does not change the fine structure constant,
the electromagnetical equations. The fundamental constants of physics vary
and our universe is steady without Big Bang.

The same method of variation of fundamental constants could be applied to
the gravitation. The assumption of decrease of the velocity of light in the
gravitational field should be made. Thus, the factor $\Omega $ is less than
1 and it depends on the gravitational potential.

We got that the quantum time rate varies but the dynamical clock rate 
remains unaltered. This discrepancy of rates can be register by a measurement 
of the light velocity with a material length standard and a quantum clock. 
The measurement will show the decrease of the light velocity with time. The 
same measurements will show that the light velocity increases in the 
gravitational field. According to general relativity a local observer 
will register the same light velocity.

\pagebreak

\end{document}